\documentclass{ws-ijmpb}

\usepackage{graphicx}
\usepackage{amssymb}

\begin{document}

\markboth{A. Sherman}
{Spin and charge fluctuations in spectra of the Hubbard model}

%
\catchline{}{}{}{}{}
%
\title{Manifestations of spin and charge fluctuations in spectra\\ of the Hubbard model}

\author{A. Sherman}

\address{Institute of Physics, University of Tartu, W. Ostwaldi Str 1, 50411 Tartu, Estonia}

\maketitle

\begin{history}
\received{Day Month Year}
\revised{Day Month Year}
\end{history}

\begin{abstract}
The influence of long-range spin and charge fluctuations on spectra of the two-dimensional fermionic Hubbard model is considered using the strong coupling diagram technique. Infinite sequences of diagrams containing ladder inserts, which describe the interaction of electrons with these fluctuations, are summed, and obtained equations are self-consistently solved for the range of Hubbard repulsions $4t\leq U\leq 8t$ and temperatures $0.3t\lesssim T\lesssim t$ with $t$ the intersite hopping constant. It was found that a metal-insulator transition curve goes from larger $U$ and $T$ to smaller values of these parameters. The temperature decrease causes the transition to the long-range antiferromagnetic order. It is responsible for the splitting out of a narrow band from a Hubbard subband with doping for $U=8t$ and low $T$. This segregated band is located near the Fermi level and forms a pseudogap here.
\end{abstract}


\section{Introduction}
The influence of charge and spin fluctuations on spectra of the fermionic Hubbard model has been attracting considerable attention due to the intimate relation of this problem to the momentum dependence of the electron self-energy and possible orderings of carriers. Short-range fluctuations were considered using Monte-Carlo simulations,\cite{Bulut} cluster approximations\cite{Kyung} and strong coupling diagram technique (SCDT).\cite{Sherman17a} More distant fluctuation were taken into account using the dynamic vertex approximation\cite{Toschi} and dual fermion approach.\cite{Rubtsov} The two latter methods use results of the dynamic mean-field approximation\cite{Georges} for calculating infinite sums of ladder diagrams. Among results of the foregoing works the description of antiferromagnetic fluctuations, a pseudogap near the Fermi level and refined boundaries of the Mott metal-insulator transition can be mentioned.

In this work, we use the SCDT\cite{Vladimir} for investigating the influence of long-range spin and charge fluctuations on spectra of the two-di\-men\-si\-o\-nal (2D) repulsive Hubbard model. In previous works\cite{Vladimir} it was shown that already two lowest-order diagrams in the expansion of the irreducible part are enough for describing the Mott metal-insulator transition. Moreover, this approximation was demonstrated\cite{Sherman15} to give spectral functions in a reasonable agreement with Monte Carlo results\cite{Preuss} for moderate $U$ and $T$. In this work, infinite sums of diagrams with ladder inserts are included into the irreducible part together with local terms of lower orders, and the obtained integral equations for the one-particle Green's function are self-consistently solved for the ranges $4t\leq U\leq 8t$ and $0.3t\lesssim T\lesssim t$. The diagrams with ladder inserts describe the interaction of electrons with spin and charge fluctuations.\cite{Sherman07} In this work, the longitudinally irreducible $ph$ vertex is approximated with its lowest-order term -- the second-order cumulant of electron operators.

Considering the case of half-filling we found that the curve separating metallic and insulating solutions is close to that obtained with account of only short-range flactuations\cite{Sherman17a} and runs from larger $U$ and $T$ to smaller values of these parameters. With account of long-range spin fluctuations the system undergoes the transition to the long-range antiferromagnetic order. The finite transition temperature, $T_{\rm AF}\approx 0.2t$, is in contradiction with the Mermin-Wagner theorem\cite{Mermin} and indicates that the used approximation somewhat overestimates the interaction.
The ordering is responsible for the splitting out of a narrow band from a Hubbard subband with doping for $U=8t$ and low $T$. The band is located near the Fermi level and form a pseudogap here.

\section{Main formulas}
The Hamiltonian of the 2D fermionic Hubbard model\cite{Hubbard} reads
\begin{equation}\label{Hamiltonian}
H=-\sum_{\bf ll'\sigma}t_{\bf ll'}a^\dagger_{\bf l'\sigma}a_{\bf l\sigma}
+\frac{U}{2}\sum_{\bf l\sigma}n_{\bf l\sigma}n_{\bf l,-\sigma},
\end{equation}
where 2D vectors ${\bf l}$ and ${\bf l'}$ label sites of a square plane lattice, $\sigma=\pm 1$ is the spin projection, $a^\dagger_{\bf l\sigma}$ and $a_{\bf l\sigma}$ are electron creation and annihilation operators, $t_{\bf ll'}$ is the hopping constant and $n_{\bf l\sigma}=a^\dagger_{\bf l\sigma}a_{\bf l\sigma}$. In this work only the nearest neighbor hopping constant $t$ is supposed to be nonzero.

\begin{figure}[t]
\centerline{\resizebox{0.7\columnwidth}{!}{\includegraphics{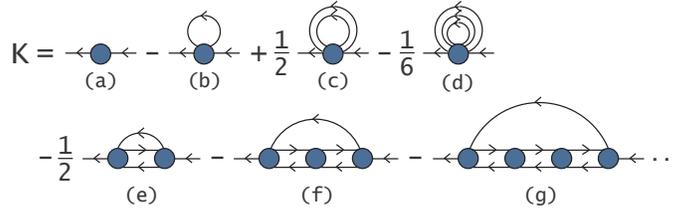}}}
\caption{Diagrams of several lowest orders in $K({\bf k}j)$.} \label{Fig1}
\end{figure}

We shall consider the electron Green's function
\begin{equation}\label{Green}
G({\bf l'\tau',l\tau})=\langle{\cal T}\bar{a}_{\bf l'\sigma}(\tau')
a_{\bf l\sigma}(\tau)\rangle,
\end{equation}
where the statistical averaging denoted by the angular brackets and time dependencies $\bar{a}_{\bf l\sigma}(\tau)=\exp{({\cal H}\tau)}a^\dagger_{\bf l\sigma}\exp{(-{\cal H}\tau)}$ are determined by the operator ${\cal H}=H-\mu\sum_{\bf l\sigma}n_{\bf l\sigma}$ with the chemical potential $\mu$. The time-ordering operator ${\cal T}$ arranges operators from right to left in ascending order of times $\tau$. In the case of strong electron correlations, $U\gg t$, for calculating this function we use the SCDT.\cite{Vladimir} In this approach, Green's function is represented by the series expansion in powers of $t_{\bf ll'}$, each term of which is a product of the hopping constants and on-site cumulants of creation and annihilation operators. These terms can be visualized as a sequence of directed lines corresponding to the hopping constants $t_{\bf ll'}$, which connect circles picturing cumulants of different orders. All these terms can be summed in the following expression for the Fourier transform of Green's function (\ref{Green}):
\begin{equation}\label{Larkin}
G({\bf k}j)=\Big\{\big[K({\bf k}j)\big]^{-1}-t_{\bf k}\Big\}^{-1},
\end{equation}
where ${\bf k}$ is the 2D wave vector, $j$ is an integer in the Matsubara frequency $\omega_j=(2j-1)\pi T$, $t_{\bf k}$ is the Fourier transform of $t_{\bf ll'}$ and $K({\bf k}j)$ is the irreducible part -- the sum of all irreducible two-leg diagrams, which cannot be divided into two disconnected parts by cutting a hopping line. Several lowest order terms of the expansion for $K({\bf k}j)$ are shown in Fig.~\ref{Fig1}.
The linked-cluster theorem is valid and partial summations are allowed in the SCDT. Therefore bare internal lines $t_{\bf k}$ in Fig.~\ref{Fig1} can be transformed into dressed ones,
\begin{equation}\label{hopping}
\theta({\bf k}j)=t_{\bf k}+t^2_{\bf k}G({\bf k}j).
\end{equation}

\begin{figure}[t]
\centerline{\resizebox{0.7\columnwidth}{!}{\includegraphics{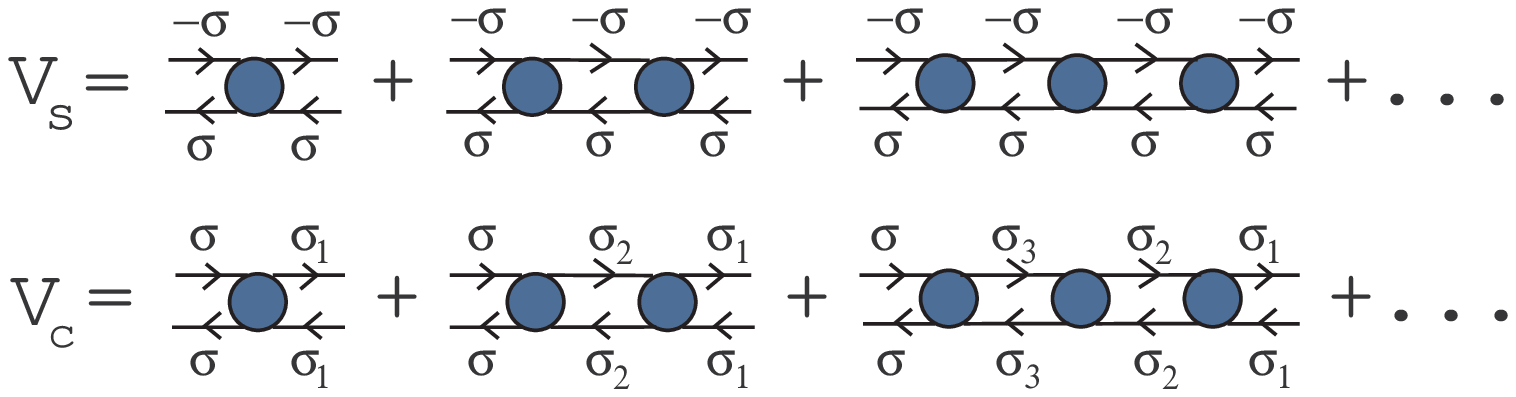}}}
\caption{Two types of infinite sums of ladders contributing to susceptibilities and, as diagram fragments, to $K({\bf k}j)$.} \label{Fig2}
\end{figure}

In Fig.~\ref{Fig1}, diagrams (a) and (b) contain on-site cumulants of the first $C_1$ and second $C_2$ orders. These diagrams as well as diagrams (c) and (d), which give small corrections, are local and their Fourier transforms are independent of momentum. In this work, in addition to diagrams (a) and (b) we take into account an infinite sequence of diagrams containing ladder inserts. Several diagrams of this type are shown in the second row in Fig.~\ref{Fig1}. These diagrams are of interest, since sums of ladders $V_c$ and $V_s$ define charge $\chi_c({\bf k}\nu)$ and spin $\chi_s({\bf k}\nu)$ susceptibilities.\cite{Sherman07} Therefore, diagrams with ladder inserts describe interactions of electrons with spin and charge fluctuations. Diagrams entering into $V_c$ and $V_s$ are shown in Fig.~\ref{Fig2}.
In the general case circles in Fig.~\ref{Fig2} denote the sum of all four-leg diagrams, which cannot be divided into two disconnected parts by cutting two horizontal particle-hole hopping lines $V_{\rm ir}^{ph}$. In this work this sum is approximated by its lowest-order term -- the second-order cumulant $C_2$. As follows from the previous results,\cite{Sherman07} it is a reasonable approximation. As a result the irreducible part reads
\begin{eqnarray}\label{K}
&&K({\bf k}j)=C_1(j)
-\frac{T}{N}\sum_{{\bf k'}j'}\theta({\bf k'}j')\big[\frac{3}{2}V_{s,\bf k- k'}(j\sigma;j\sigma;j',-\sigma;j',-\sigma)\nonumber\\
&&\quad\quad+\frac{1}{2}V_{c,\bf k-k'}(j\sigma;j\sigma;j'\sigma;j'\sigma)\big]
+\frac{T^2}{2N^2}\sum_{{\bf k'}j'\nu}\theta({\bf k'}j'){\cal T}_{\bf k-k'}(j+\nu,j'+\nu) \nonumber\\
&&\quad\quad\times\Big[C_2(j\sigma;j+\nu,\sigma;j'+\nu,-\sigma;j',-\sigma)
C_2(j+\nu,\sigma;j\sigma;j',-\sigma;j'+\nu,-\sigma)\nonumber\\
&&\quad\quad+\sum_{\sigma'} C_2(j\sigma;j+\nu,\sigma';j'+\nu,\sigma';j'\sigma)
C_2(j+\nu,\sigma';j\sigma;j'\sigma;j'+\nu,\sigma')\Big],
\end{eqnarray}
where sums of ladder diagrams $V_s$ and $V_c$ satisfy Bethe-Salpeter equations (BSE) and ${\cal T}_{\bf k}(jj')=N^{-1}\sum_{\bf k'}\theta({\bf k+k'},j)\theta({\bf k'}j')$.

Equations for second-order cumulants were derived in earlier works.\cite{Vladimir,Sherman07} They are rather cumbersome. However, the equations can be significantly simplified in the case
\begin{equation}\label{condition}
T\ll\mu,\quad T\ll U-\mu.
\end{equation}
For $U\gg T$ this range of $\mu$ contains the most interesting cases of half-filling, $\mu=U/2$, and moderate doping. For the conditions (\ref{condition}) the first- and second-order cumulants read
\begin{eqnarray}
&&C_1(j)=\frac{1}{2}\left[g_1(j)+g_2(j)\right],\nonumber\\
&&C_2(j+\nu,\sigma;j\sigma';j'\sigma';j'+\nu,\sigma)=\frac{1}{4T}\big[\delta_{jj'} \big(1-2 \delta_{\sigma\sigma'}\big)\label{cumulants}\\
&&\quad+\delta_{\nu 0}\big(2-\delta_{\sigma\sigma'}\big)\big]
a_1(j'+\nu)a_1(j)-\delta_{\sigma,-\sigma'}B(jj'\nu),\nonumber
\end{eqnarray}
where
\begin{eqnarray*}
&&g_1(j)=(i\omega_j+\mu)^{-1},\quad g_2(j)=(i\omega_j+\mu-U)^{-1}, \\
&&B(jj'\nu)=\frac{1}{2}\big[a_1(j'+\nu)a_2(jj')+a_4(j'+\nu,j+\nu)a_3(jj')\\
&&\quad\quad+a_2(j'+\nu,j+\nu)a_1(j)+a_3(j'+\nu,j+\nu)a_4(jj')\big], \\
&&a_1(j)=g_1(j)-g_2(j),\quad a_2(jj')=g_1(j)g_1(j'),\\
&&a_3(jj')=g_2(j)-g_1(j'),\quad a_4(jj')=a_1(j)g_2(j').
\end{eqnarray*}

Substituting (\ref{cumulants}) into the BSE for $V_s$ we get
\begin{eqnarray}\label{Vsnew}
&&V_{s{\bf k}}(j+\nu,j,j',j'+\nu)=\frac{1}{2}f_1({\bf k},j+\nu,j'+\nu)\nonumber\\
&&\quad\times\bigg\{2C_2(j+\nu,\sigma;j\sigma;j',-\sigma;j'+\nu,-\sigma)
+\bigg[a_2(j'+\nu,j+\nu)\nonumber\\
&&\quad-\frac{\delta_{jj'}}{T}a_1(j'+\nu)\bigg]y_1({\bf k}jj')
+a_1(j'+\nu)y_2({\bf k}jj')\nonumber\\
&&\quad+a_4(j'+\nu,j+\nu)y_3({\bf k}jj')+a_3(j'+\nu,j+\nu)y_4({\bf k}jj')\bigg\},
\end{eqnarray}
where $f_1({\bf k}jj')=\bigg[1+\frac{1}{4}a_1(j)a_1(j'){\cal T}_{\bf k}(jj')\bigg]^{-1}$ and
$$y_i({\bf k}jj')=T\sum_\nu a_i(j+\nu,j'+\nu){\cal T}_{\bf k}(j+\nu,j'+\nu)
V_{s{\bf k}}(j+\nu,j,j',j'+\nu).$$
Equations for $y_i({\bf k}jj')$ are derived from Eq.~(\ref{Vsnew}),
\begin{eqnarray}\label{eq_for_y}
&&y_i({\bf k}jj')=b_i({\bf k}jj')+\bigg[c_{i2}({\bf k}jj')-\frac{\delta_{jj'}}{T}c_{i1}({\bf k}jj')\bigg]y_1({\bf k}jj')\nonumber\\
&&\quad+c_{i1}({\bf k}jj')y_2({\bf k}jj')+c_{i4}({\bf k}jj')y_3({\bf k}jj')
+c_{i3}({\bf k}jj')y_4({\bf k}jj'),
\end{eqnarray}
where
\begin{eqnarray*}
&&b_i({\bf k}jj')=-\frac{1}{4}a_i(jj')a_1(j)a_1(j'){\cal T}_{\bf k}(jj')f_1({\bf k}jj')
+\bigg[a_2(jj')-\frac{\delta_{jj'}}{T}a_1(j)\bigg]c_{i1}({\bf k}jj')\\
&&\quad+a_1(j)c_{i2}({\bf k}jj')+a_4(jj')c_{i3}({\bf k}jj')+a_3(jj')c_{i4}({\bf k}jj'),\\
&&c_{ii'}({\bf k}jj')=\frac{T}{2}\sum_\nu a_i(j+\nu,j'+\nu)a_{i'}(j'+\nu,j+\nu)\\
&&\quad\times{\cal T}_{\bf k}(j+\nu,j'+\nu)f_1({\bf k},j+\nu,j'+\nu).
\end{eqnarray*}
Thus, the solution of the BSE was reduced to the solution of the system of four linear equations (\ref{eq_for_y}) with respect to four variables $y_i({\bf k}jj')$.

In the same manner we can solve the BSE for $V_c$,
\begin{eqnarray}\label{Vcnew}
&&V_{c\bf k}(j+\nu,j,j',j'+\nu)=\frac{1}{2}f_2({\bf k},j+\nu,j'+\nu)\Big[2\sum_{\sigma'}C_2(j+\nu,\sigma';j\sigma;j'\sigma;j'+\nu,\sigma')\nonumber\\
&&\quad-a_2(j'+\nu,j+\nu)z_1({\bf k}jj')-a_1(j'+\nu)z_2({\bf k}jj')-a_4(j'+\nu,j+\nu)z_3({\bf k}jj')\nonumber\\
&&\quad-a_3(j'+\nu,j+\nu)z_4({\bf k}jj')\Big],
\end{eqnarray}
where $f_2({\bf k}jj')=\bigg[1-\frac{3}{4}a_1(j)a_1(j'){\cal T}_{\bf k}(jj')\bigg]^{-1}$ and
$$z_i({\bf k}jj')=T\sum_\nu a_i(j+\nu,j'+\nu){\cal T}_{\bf k}(j+\nu,j'+\nu)
V_{c\bf k}(j+\nu,j,j',j'+\nu).$$
The four quantities $z_i({\bf k}jj')$ are obtained from the system of linear equations, \begin{eqnarray}\label{eq_for_z}
&&z_i({\bf k}jj')=d_i({\bf k}jj')-e_{i2}({\bf k}jj')z_1({\bf k}jj')
-e_{i1}({\bf k}jj')z_2({\bf k}jj')\nonumber\\
&&\quad-e_{i4}({\bf k}jj')z_3({\bf k}jj')-e_{i3}({\bf k}jj')z_4({\bf k}jj')
\end{eqnarray}
with
\begin{eqnarray*}
&&d_i({\bf k}jj')=\frac{3}{4}a_i(jj')a_1(j)a_1(j'){\cal T}_{\bf k}(jj')f_2({\bf k}jj')
-a_2(jj')e_{i1}({\bf k}jj')\\
&&\quad-a_1(j)e_{i2}({\bf k}jj')
-a_4(jj')e_{i3}({\bf k}jj')-a_3(jj')e_{i4}({\bf k}jj'),\\
&&e_{ii'}({\bf k}jj')=\frac{T}{2}\sum_\nu a_i(j+\nu,j'+\nu)a_{i'}(j'+\nu,j+\nu)\\
&&\quad\times{\cal T}_{\bf k}(j+\nu,j'+\nu)f_2({\bf k},j+\nu,j'+\nu).
\end{eqnarray*}

Equations~(\ref{Larkin}), (\ref{K}), (\ref{eq_for_y}) and (\ref{eq_for_z}) form a closed set for calculating $G({\bf k}j)$ that can be solved by iteration. Green's function of the Hubbard-I approximation\cite{Hubbard} was chosen as the starting one in this procedure. Calculations were mainly performed in 8$\times$8 and 16$\times$16 lattices.

\section{Results of calculations}
\begin{figure}[t]
\centerline{\resizebox{0.5\columnwidth}{!}{\includegraphics{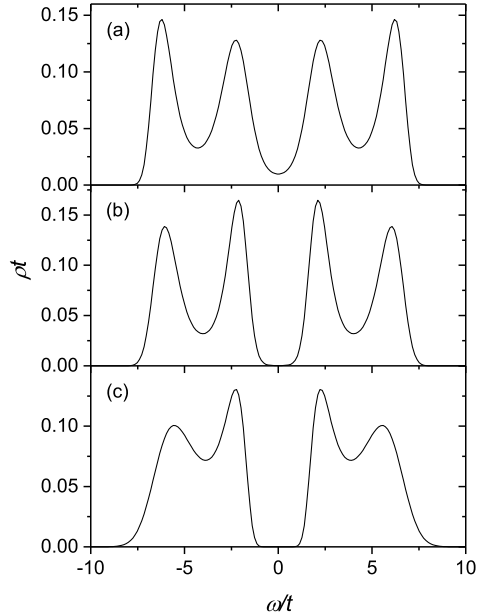}}}
\caption{The density of states for $U=8t$, $\mu=4t$ and temperatures $T=1.07t$ (a), $0.64t$ (b) and $0.32t$ (c).} \label{Fig3}
\end{figure}
In this sections calculated densities of states (DOS) $\rho(\omega)=-(\pi N)^{-1}\sum_{\bf k}{\rm Im}G({\bf k}\omega)$ are shown. The analytic continuation from imaginary to real frequencies $\omega$ was performed using the maximum entropy method.\cite{Jarrell} Let us start from the case of half-filling, $\mu=U/2$. Figure~\ref{Fig3} shows the temperature variation of the DOS for $U=8t$. Mott gaps around $\omega=0$ are well seen in panels (b) and (c), pointing to insulating states. The DOS in panel (a) has a small finite intensity near $\omega=0$. This finite intensity may be an artefact of the analytic continuation. As in the lower-order calculations,\cite{Sherman17a,Sherman15} the density is suppressed near frequencies $\omega=\pm U/2$. These pseudogaps together with the dip near $\omega=0$ give a four-band shape to the DOS. An analogous four-band structure was observed in Monte Carlo simulations.\cite{Preuss} The pseudogaps were related\cite{Sherman15} to multiple reabsorption of electrons with the creation of states with double site occupancies.

\begin{figure}[t]
\centerline{\resizebox{0.5\columnwidth}{!}{\includegraphics{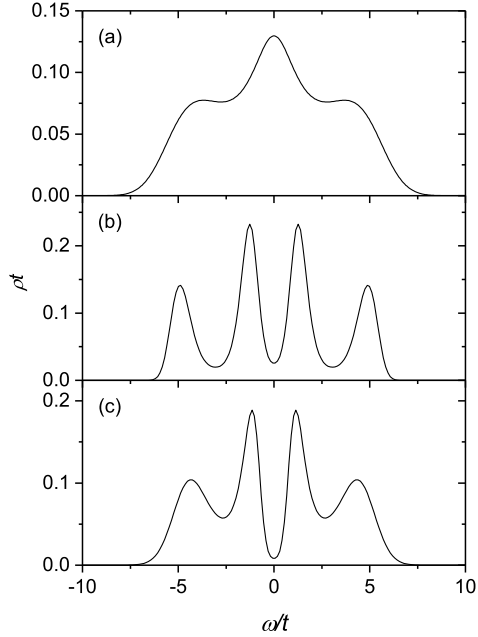}}}
\caption{The density of states for $U=5.1t$, $\mu=2.55t$ and temperatures $T=0.62t$ (a), $0.49t$ (b) and $0.29t$ (c).} \label{Fig4}
\end{figure}

The temperature variation of the DOS for somewhat smaller repulsion, $U=5.1t$, is shown in Fig.~\ref{Fig4}. A quasiparticle peak at $\omega=0$ is seen in panel (a). This spectral feature is inherent in a metallic state. With a small temperature decrease the peak is changed to a dip or a gap characteristic for an insulator (panel b). For this $U$ the metal-insulator transition occurs near $T=0.6t$, which is close to the result obtained with account of only short-range fluctuations.\cite{Sherman17a} For other considered values of $U$ transitions temperatures were found also to be close to those obtained in Ref.~[\refcite{Sherman17a}]. It can be concluded that short-range fluctuations play the main role in determining the locations of the transition curve in the considered parameter ranges.
The curve is shown in Fig.~7 in Ref.~[\refcite{Sherman17a}]. It goes from larger $U$ and $T$ to smaller values of these quantities.

The above results were obtained for temperatures $T\gtrsim 0.3t$. At a further temperature lowering the convergence of the iteration procedure is impaired. This is connected with the decrease of the determinant of the system of four linear equations (\ref{eq_for_y}). The decrease is most pronounced at ${\bf k}=(\pi,\pi)$ and $j=j'$. The vanishing determinant leads to the divergence of quantities $y_i$ that entails the divergence of the ladder sum $V_{s\bf k}$ (\ref{Vsnew}) and spin susceptibility $\chi_s({\bf k}\nu)$ at the antiferromagnetic ordering vector ${\bf k}=(\pi,\pi)$ and at the frequency $\omega_\nu=2\nu\pi T=0$. Hence the vanishing determinant signals the transition to the long-range antiferromagnetic order. The transition temperature $T_{\rm AF}$ is nonzero, for $U=8t$ it is approximately equal to $0.24t$. Transition temperatures for other considered values of $U$ in 8$\times$8 and 16$\times$16 lattices are close to $T_{\rm AF}$ specified above. The finite value of this quantity is in contradiction with the Mermin-Wagner theorem.\cite{Mermin} This indicates that the approximation of the four-leg diagram $V^{ph}_{\rm ir}$ by the second-order cumulant, which was used above, somewhat overestimates the interaction.

The determinant of the second system of linear equations (\ref{eq_for_z}) decreases also with temperature. However, this decrease is much smaller than that in the system (\ref{eq_for_y}), and the determinant never goes to zero. Therefore, the ladder sum $V_{c\bf k}$ (\ref{Vcnew}) and charge susceptibility $\chi_c({\bf k}\nu)$ do not diverge. This result indicates that there is no charge ordering in the normal-state $t$-$U$ Hubbard model. A deviation from half-filling does not change this conclusion.

\begin{figure}[t]
\centerline{\resizebox{0.5\columnwidth}{!}{\includegraphics{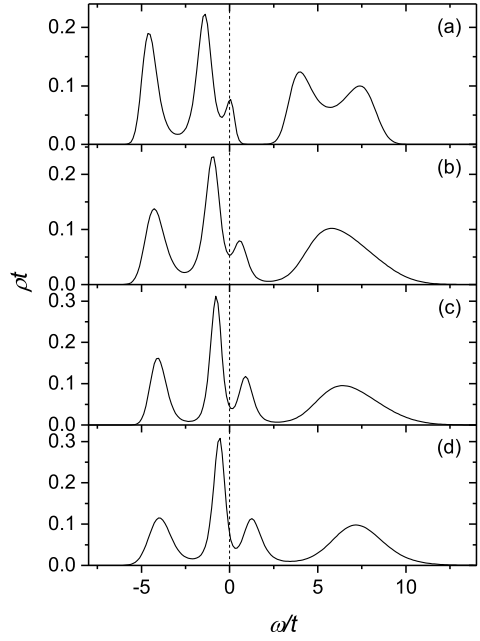}}}
\caption{The density of states for $U=8t$, $T=0.32t$, $\mu=2.5t$ (a, $\bar{n}=0.97$), $\mu=2t$ (b, $\bar{n}=0.93$), $\mu=1.7t$ (c, $\bar{n}=0.89$) and $\mu=1.5t$ (d, $\bar{n}=0.86$).} \label{Fig5}
\end{figure}
Let us consider changes in the DOS caused by this deviation. Due to the particle-hole symmetry of Hamiltonian (\ref{Hamiltonian}) only the case $\bar{n}<1$ will be examined, where $\bar{n}=2\int_{-\infty}^{\infty}\rho(\omega)[\exp{(\omega/T)}+1]^{-1}d\omega$ is the electron concentration. The case $U=8t$ and $T=0.32t$ is shown in Fig.~\ref{Fig5}. The most prominent consequence of doping is the splitting out of a narrow band from the Hubbard subband, in which the Fermi level is located. Between the narrow band and the subband a pseudogap is formed near $\omega=0$. Notice that the segregated band and the pseudogap are only observed for large repulsions and temperatures close to $T_{\rm AF}$, which points to a key role of long-range antiferromagnetic fluctuations in their formation. This resembles the mechanism of the pseudogap formation in the 2D $t$-$J$ model, where the pseudogap arises owing to the spin-polaron band, which is segregated from a Hubbard subband due to an interaction of carriers with spin excitations in the antiferromagnetic background.\cite{Sherman97} As seen from comparison of Figs.~\ref{Fig3}(c) and \ref{Fig5}, the appearance of the segregated band is accompanied with a strong redistribution of the spectral intensity, which is an indication of strong electron correlations. Doping leads to much weaker spectral redistributions in cases of larger temperatures and smaller repulsions.

With doping the determinant $\Delta$ of the system (\ref{eq_for_y}) remains at a minimum for ${\bf k}=(\pi,\pi)$ and $j=j'$, and this value increases with a rise of $|1-\bar{n}|$, that points to a reduction of the antiferromagnetic correlation length with doping. An analogous behaviour is observed in cuprates.\cite{Birgeneau}

\section{Concluding remarks}
In this work, the SCDT was used for investigating the influence of long-range spin and charge fluctuations on electron spectra of the two-di\-men\-si\-o\-nal $t$-$U$ Hubbard model. The infinite sequence of diagrams with ladder inserts, which were constructed from cumulants of the first and second orders, was included into the irreducible part. These diagrams give an account of the interactions of electrons with spin and charge fluctuations. The obtained equations were solved by iteration, mainly in 8$\times$8 and 16$\times$16 lattices, for the ranges of Hubbard repulsions $4t\leq U\leq 8t$ and temperatures $0.3t\lesssim T\lesssim t$.

At half-filling, the inclusion of long-range fluctuations does not cardinally change the location of the metal-insulator transition curve in the $U$-$T$ plane in comparison with that obtained with account of only short-range fluctuations. In the used approximation it is these latter fluctuations that determine the location of the curve in the considered parameter ranges. The curve goes from larger $U$ and $T$ to smaller values of these quantities.

The inclusion of long-range spin fluctuations leads to the transition to the long-range antiferromagnetic order at $T_{\rm AF}\approx 0.2t$. The non-zero value of $T_{\rm AF}$ indicates that the used approximation for the four-leg vertex somewhat overestimates the interaction. The used approach allows one to improve this result by including sums of transversal $ph$ and $pp$ ladders in the vertex $V_{\rm ir}^{ph}$. With doping the deterioration of the antiferromagnetic ordering is observed, which is analogous to that in cuprate perovskites. We found no indication of a charge ordering in the considered model.

For strong repulsions and low temperatures doping leads to the splitting out of the narrow band from a Hubbard subband. Between these band and subband a pseudogap is formed near the Fermi level. As in the 2D $t$-$J$ model, the segregated band and the pseudogap arise due to the interaction of electrons with long-range antiferromagnetic fluctuations.

\end{document}